# *BFL*: a Logic to Reason about Fault Trees

Stefano M. Nicoletti* E. Moritz Hahn* Mariëlle Stoelinga*†
*University of Twente, Formal Methods and Tools, Enschede, the Netherlands.
†Radboud University, Department of Software Science, Nijmegen, the Netherlands.
{s.m.nicoletti,e.m.hahn,m.i.a.stoelinga}@utwente.nl

***Abstract*—Safety-critical infrastructures must operate safely and reliably. Fault tree analysis is a widespread method used to assess risks in these systems: fault trees (FTs) are required — among others — by the Federal Aviation Authority, the Nuclear Regulatory Commission, in the ISO26262 standard for autonomous driving and for software development in aerospace systems. Although popular both in industry and academia, FTs lack a systematic way to formulate powerful and understandable analysis queries. In this paper, we aim to fill this gap and introduce Boolean Fault tree Logic (*BFL*), a logic to reason about FTs. *BFL* is a simple, yet expressive logic that supports easier formulation of complex scenarios and specification of FT properties. Alongside *BFL*, we present model checking algorithms based on binary decision diagrams (BDDs) to analyse specified properties in *BFL*, patterns and an algorithm to construct counterexamples. Finally, we propose a case-study application of *BFL* by analysing a COVID19-related FT.**

## I. Introduction

Our self-driving cars, power plants, and transportation systems must operate in a safe and reliable way. Risk assessment is a key activity to identify, analyze and prioritize the risk in a system, and come up with (cost-)effective countermeasures.

Fault tree analysis (FTA) [1, 2] is a widespread formalism to support risk assessment. FTA is applied to many safety-critical systems and the use of fault trees is required for instance by the Federal Aviation Authority (FAA), the Nuclear Regulatory Commission (NRC), in the ISO 26262 standard [3] for autonomous driving and for software development in aerospace systems. A fault tree (FT) models how component failures arise and propagate through the system, eventually leading to system level failures. Leaves in a FT represent *basic events* (BEs), i.e. elements of the tree that need not be further refined. Once these fail, the failure is propagated through the *intermediate events* (IEs) via *gates*, to eventually reach the *top level event* (TLE), which symbolizes system failure. In the (sub)tree represented in Fig. 1, the TLE— *Existence of COVID-19 Pathogens/Reservoir* — is refined by an OR-gate (*CP/R*). For *CP/R* to fail, either pathogens must exist on the workplace, i.e., *Existence of COVID-19 Pathogens* (*CP*), or there must be an infected object of some kind, i.e., *Existence of COVID-19 Reservoir* (*CR*) has to happen. Both *CP* and *CR* are AND-gates: for them to fail, all their respective children need to fail. For *CP* this means that an *Infected worker joining the team* (*IW*) and a failure in detecting this, i.e., *Detection error* (*H3*) must happen. For *CR* this means that an *Infected object used by the team* (*IT*) and a *General disinfection error* (*H2*) must happen. Fault tree analysis supports qualitative and quantitative analysis. Qualitative analysis aims at pointing out root causes and critical paths in the system. Typically, one identifies the *minimal cut sets* (MCSs) of a FT, i.e. minimal sets of BEs that, when failed, cause the system to fail. One can also identify *minimal path sets* (MPSs), i.e. minimal sets of BEs that - when operational - guarantee that the system will remain operational. Quantitative analysis allows to compute relevant dependability metrics, such as the system reliability, availability and mean time to failure. A formal background on FTs is provided in Sec. II.

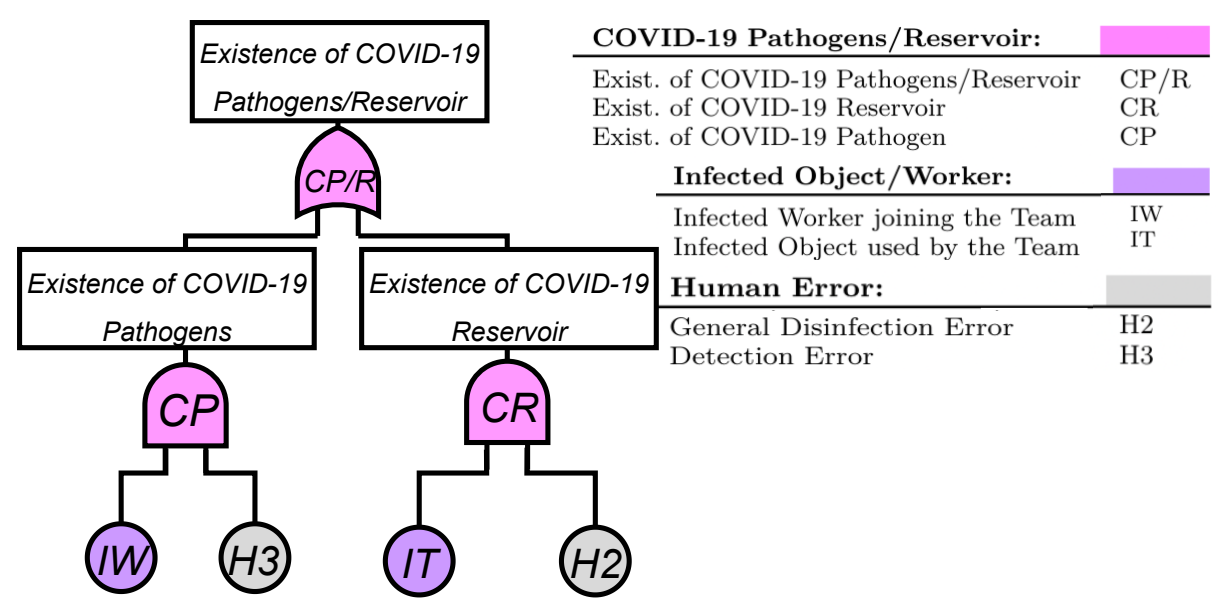

Figure 1: A simple FT (excerpt from Fig. 2).

In spite of their popularity, FTs lack a systematic way to formulate powerful yet understandable analysis queries. The qualitative and quantitative analysis questions mentioned above are formulated by ad hoc means. In particular, if scenarios are analysed, the fault tree has to be altered, for instance if one likes to compute the system reliability given that certain subsystems have failed.

**Boolean Fault tree Logic:** In this paper, we aim to fill this gap and introduce *BFL*, a Boolean logic to reason about FTs. *BFL* is based on concrete insights and needs gathered through series of questions targeted at a FT practitioner from industry [4]. In this paper, we aim to fill this gap and introduce *BFL*, a Boolean logic to reason about FTs. *BFL* is based on concrete insights and needs

*This work was partially funded by the NWO grant NWA.1160.18.238 (PrimaVera), and the European Union's Horizon 2020 research and innovation programme under the Marie Skłodowska-Curie grant agreement No 101008233, and the ERC Consolidator Grant 864075 (*CAESAR*).

gathered through series of questions targeted at a FT practitioner from industry [4]. The atomic propositions in this logic are the FT elements, i.e., both the BEs and the IEs. As usual, formulae can be combined through Boolean connectives. Furthermore, we include operators for setting evidence, and for MCSs and MPSs. In this way, we obtain a simple, yet expressive logic to reason about FTs that supports easier formulation of scenarios. Among others:

- We can set evidence to analyse what-if scenarios. E.g., what are the MCSs, given that BE $A$ or subsystem $B$ has failed? What are the MPSs given that $A$ or $B$ have not failed?
- We can check whether two elements are independent or if they share a children that can influence their status.
- We can check whether the failure of one (or more) element $E$ always leads to the failure of TLE.
- We can set upper/lower boundaries for failed elements. E.g., would element $E$ always fail if at most/at least two out of $A$, $B$ and $C$ were to fail?

Moreover, if a property does not hold, *BFL* allows us to generate counterexamples, to show why the property fails. E.g., if some set $S$ is not a MCS, we can generate additional BEs $e_1, ..e_n$ that need to fail in order for $S$ to be a MCS.

**Model checking:** As a first step, we focus on non-probabilistic logic to set a robust baseline that can be easily extended, focusing on useful algorithms and counterexample generation. We present algorithms to answer two model checking queries and we discuss a third scenario. First, we propose an algorithm to check if $T, \overline{b} \models \chi$ holds, i.e. if a *BFL* formula $\chi$ holds for a given FT $T$ and a *status vector* $\overline{b}$. The latter indicates, for each BE in $T$, whether that BE has failed. Second, we present a procedure that computes all status vectors $[\![\overline{b}]\!]$ for which $T, \overline{b} \models \chi$ holds. Finally, given $\overline{b}$ and $\chi$, we reflect on procedures to synthesize, if it exists, a FT $T$ such that $T, \overline{b} \models \chi$ holds. The algorithms exploit clever manipulations of Binary Decision Diagrams (BDDs). In order to translate formulae to BDDs, we identify FT elements that appear in a given formula. We then construct BDDs only for these elements and store the resulting BDDs in case they are needed in further computations. Finally, we manipulate these BDDs to reflect the semantics of the operators in *BFL*. Once the BDD for the formula is obtained, we either walk down from the root node following truth assignments given in a specific vector $\overline{b}$ or — if no vector is given — we collect every path that leads to the terminal 1 to compute all satisfying vectors $[\![\overline{b}]\!]$ for that formula. BDDs are naturally applicable to FTs, since FTs are essentially Boolean functions and BDDs provide compact representations of Boolean functions. Thus, BDDs are heavily exploited in fault tree analysis [1, 5, 6].

**Counterexamples:** Moreover, we provide an algorithm and several patterns to construct counterexamples. We showcase them by selecting some example formulae and we represent resulting counterexamples. Given a status vector $\overline{b}$ and a formula $\chi$, if $\overline{b}$ does not satisfy $\chi$ we compute a new $\overline{b'}$ such that $\overline{b'} \models \chi$ for the given $T$. We then represent how failures propagate through $T$ in light of the newly found vector.

**Contributions:** To summarize, in this work:

1) We develop a logic for FTs that enables the construction of complex queries, to express numerous relevant scenarios.
2) We provide model checking algorithms to check properties defined in the logic.
3) We present an algorithm and patterns to construct and represent useful counterexamples.
4) We showcase the potential of our logic by applying it to a medium-sized COVID-19 related example.

**Related work:** Numerous well-known logics describe properties of state-transition systems, such as labelled transition systems and Markov models. Examples comprise CTL [7], LTL [8], and their variants for Markov models, PCTL [9] and PLTL [10]. State-transition systems are usually not written by hand, but result as the semantics of high-level description mechanisms, such as AADL [11], the hardware description language VHDL [12] or model description languages such as JANI [13] or PRISM [14]. These logics are thus not used to reason about the structure of such models (e.g. the placement of circuit elements in a VHDL model or the structure of modules in a PRISM model), but on the temporal behaviour of the underlying state-transition system. Similarly, related work on model checking on FTs [15, 16, 17, 18] exhibits significant differences: these works perform model checking by referring to states in the underlying Markov model, not to events in the given FT. We develop a logic to express properties over FTs and present counterexamples generation methods on the level of FTs. Since existing counterexample generation methods [19] works on the underlying Markov models, it is difficult to relate these to the FT. Another related line of work are software specification languages, such as the Java Modelling Language (JML) [20]. JML uses Hoare style pre- and postconditions and invariants, to prove the correctness of programs using the design by contract paradigm. Logics operating on the structure of high-level description mechanisms (other than FTs) exist. The Object Constraint Language (OCL) for UML is a textual query and description language developed by IBM [21]. [22] uses description logic (DL) to detect and resolve consistency conflicts between (different versions of) UML diagrams. [23] is involved with the automatic analysis of UML class diagrams using DL. In [24], the author provides a formulation of *Pandora*, a logic for the qualitative analysis of temporal FTs. In spite of the use of logic to capture FT properties, [24] focuses on the analysis of time, introducing gates that are different from the ones considered in this work: the Priority-AND-gate (PAND), the Simultaneous-AND-gate (SAND), and the Priority-OR gate (POR). In *BFL* we do not (yet) consider time and we focus on AND, OR and VOT-gates in order to create a flexible yet powerful logic that can be further extended in the future. Furthermore, [24] focuses more on the algorithmic part of FTA while leaving out any formalization of FTs or the logic defined upon. To the best of our knowledge, no other work considers the option of developing a logic specific to FTs. Literature related to FTs and BDDs is opportunely referenced and contextualized in Sec. II and Sec. V.

**Structure of the paper:** Sec. II provides background on FTs, Sec. III showcases the logic, Sec. IV shows how FT properties can be specified in *BFL*, Sec. V presents our model checking algorithms, Sec. VI constructs and represents useful counterexamples, Sec. VII shows an application of our logic to a medium-sized case study and Sec. VIII concludes the paper and reflects on future work.

## II. Fault Trees: Background

Developed in the early '60s [25], FTs are directed acyclic graphs (DAGs) that model how low-level failures can propagate and cause a system-level failure. The overall failure of a system is captured by a *top level event* (TLE), that is refined through the use of *gates*. FTs come with different gate types. For the purposes of our paper, we will focus on *static* fault trees, featuring OR-gates, AND-gates and VOT*(k/N)*-gates. In order for a low-level failure to propagate, at least one child of an OR-gate has to fail, all the children of an AND-gate must fail, and at least $k$ out of $N$ elements must fail for a $\texttt{VOT}(k/N)$-gate to fail. When gates can no longer be refined, we reach the *basic events* (BEs) which are the leaves of the tree. FTs enable both qualitative and quantitative analyses. On the qualitative side, *minimal cut sets* (MCSs) and *minimal path sets* (MPSs) have a paramount role in highlighting root causes of failures and critical paths in the system. MCSs are minimal sets of events that - when failed - cause the failure of the TLE. MPSs are minimal sets of events that - when remaining operational - guarantee that the TLE will remain operational.

**Definition 1** (*Fault Tree*)**.** A *Fault Tree* is a 4-tuple $T = \langle \texttt{BE}, \texttt{IE}, t, ch \rangle$ consisting of the following components:

- $\texttt{BE}$ is the set of basic events.
- $\texttt{IE}$ is the set of intermediate elements with $\texttt{BE} \cap \texttt{IE} = \emptyset$. Let $e_{top} \in \texttt{IE}$ be the top element of the fault tree.
- $\texttt{E} = \texttt{BE} \cup \texttt{IE}$ is the set of all the elements.
- $t : \texttt{IE} \rightarrow \mathit{GateTypes}$ is a function that maps each intermediate element to its gate type, with $\mathit{GateTypes} = \{\texttt{AND}, \texttt{OR}\}$.
- $ch : \texttt{IE} \rightarrow \mathscr{P}(\texttt{E})\backslash\emptyset$ is a function that maps each intermediate element to its children (its inputs). With $e \in \texttt{IE}$, we require that $ch(e) \neq \emptyset$.

In order to be meaningful, FTs have to meet the following well-formedeness condition: the graph formed by $\mathscr{G} = \langle \texttt{E}, ch \rangle$ must be acyclic with a unique root in $e_{top}$ that is reachable from all other nodes.

We can extend *GateTypes* with any gate derived from AND and OR-gates. E.g., we can add $\texttt{VOT}(k/N)$ by extending *GateTypes* as follows: $\mathit{GateTypes} = \{\texttt{AND}, \texttt{OR}\} \cup \{\texttt{VOT}(k/N) \mid k, N \in \mathbb{N}^{>1}, k \leq N\}$. Furthermore, it suffices to require that $|ch(e)| = N$ if $t(e) = \texttt{VOT}(k/N)$. The behaviour of a FT $T$ can be rigorously expressed through its *structure function* [1] - $\Phi_T$: if we assume the convention that a BE has value 1 if failed and 0 if operational, the structure function indicates the status of the TLE given the status of all the BEs of $T$.

**Definition 2** (*Structure Function*)**.** The structure function of a fault tree $T$ is a function $\Phi_T \colon \mathbb{B}^n \times \texttt{E} \rightarrow \mathbb{B}$ that takes as input a status vector $\overline{b} = (b_1, \ldots, b_k)$ of $k$ Booleans, where $b_i = 1$ if the i-th BE has failed and $b_i = 0$ if it is operational, and an arbitrary element $e \in \texttt{E}$.

$$\Phi_T(\overline{b}, e) = \begin{cases} b_i & \text{if } e = e_i \in \texttt{BE} \\ \bigvee_{e' \in ch(e)} \Phi_T(\overline{b}, e') & \text{if } e \in \texttt{IE} \text{ and } t(e){=}\texttt{OR} \\ \bigwedge_{e' \in ch(e)} \Phi_T(\overline{b}, e') & \text{if } e \in \texttt{IE} \text{ and } t(e){=}\texttt{AND} \end{cases}$$

Thus, for each set of BEs we can identify its characteristic vector $\overline{b}$. Note that the semantics for $e \in \texttt{IE}$ and $t(e){=}\texttt{VOT}(k/N)$ is given by

$$\left( \sum_{e' \in ch(e)} \Phi_T(\overline{b}, e') \right) \geq k$$

Next we define the classical notions of minimal cut sets and minimal path sets [1]. A cut set is any set of basic events that causes the TLE to occur, i.e., for which the structure function evaluates to 1. A path set is any set of basic events that does not cause the TLE to occur, i.e., for which the structure function evaluates to 0.

**Definition 3.** A status vector $\overline{b}$ is a *cut set (CS)* for element $e \in \texttt{E}$ of a given tree $T$ if $\Phi_T(\overline{b}, e) = 1$. A *minimal cut set (MCS)* is a cut set of which no subset is a cut set. Formally: $\overline{b}$ is a MCS for $e \in \texttt{E}$ of $T$ if $\Phi_T(\overline{b}, e) = 1 \wedge \forall \overline{b'} \subset \overline{b}, \Phi_T(\overline{b}, e) = 0$.

**Definition 4.** A status vector $\overline{b}$ is a *path set (PS)* for element $e \in \texttt{E}$ of a given tree $T$ if $\Phi_T(\overline{b}, e) = 0$. A *minimal path set (MPS)* is a path set of which no subset is a path set. Formally: $\overline{b}$ is a MPS for $e \in \texttt{E}$ of $T$ if $\Phi_T(\overline{b}, e) = 0 \wedge \forall \overline{b'} \subset \overline{b}, \Phi_T(\overline{b}, e) = 1$.

The FT in Fig. 1 has two minimal cut sets: $\{\mathit{IW}, \mathit{H3}\}$ and $\{\mathit{IT}, \mathit{H2}\}$. Its minimal paths sets are $\{\mathit{IW}, \mathit{IT}\}$, $\{\mathit{IW}, \mathit{H2}\}$, $\{\mathit{H3}, \mathit{IT}\}$, and $\{\mathit{H3}, \mathit{H2}\}$.

## III. A Logic to Reason about FTs

### A. Syntax

Below, we present the syntax of *BFL*. The atomic propositions in *BFL* can be any FT element $e$. As usual,

formulae can be combined through Boolean connectives. Furthermore, we can set evidence: $\phi[e \mapsto 0]$ sets the element $e$ to 0, and $\phi[e \mapsto 1]$ sets $e$ to 1. Finally, *BFL* allows reasoning about MCSs.

We construct our logic on two syntactic layers, represented with $\phi$ and $\psi$. With $\chi$ we will refer to any formula in either the first or the second layer. Formulae in the first layer are evaluated on a single status vector, while formulae in the second layer allow quantification over status vectors. Further, the IDP operator expresses whether two formulae are independent, i.e. whether there is no event which influences the truth value of both formulae.

$$\begin{aligned}\phi &::= e \mid \neg\phi \mid \phi \wedge \phi \mid \phi[e \mapsto 0] \mid \phi[e \mapsto 1] \mid \mathrm{MCS}(\phi)\\ \psi &::= \exists\phi \mid \forall\phi \mid \mathrm{IDP}(\phi,\phi)\end{aligned}$$

Note that $\phi[e \mapsto 0]$ is not equivalent to $\phi \wedge \neg e$. Taking $\phi$ equal to $\neg e$, we will have that $(\neg e)[e \mapsto 0]$ evaluates to true, while $(\neg e) \wedge \neg e$ is equivalent to $\neg e$. Moreover, we have operators to express MCSs for a given formula: in addition to MCSs for TLE, we can evaluate whether a given status vector is a MCS for an intermediate event (recall Def. 3). Lastly, the second layer provides us with the possibility to quantify over formulae and to check if two formulae (e.g., two intermediate events) are independent. We consider two formulae to be independent whenever they do not share any *influencing basic event* (IBE) i.e., any event the value of which influences the truth value of the given formulae.

**Syntactic sugar:** We define several derived operators. Apart from extra Boolean operators, MPSs denotes the minimal path sets, and $\mathrm{SUP}(e)$ means that event $e \in \mathtt{E}$ is superfluous, i.e., its value does not influence the TLE. Finally, the voting operator $\underset{\geq k}{\mathrm{Vot}}(\phi_1, \ldots, \phi_N)$ holds if at least $k$ of the formulae $\phi_1, \ldots, \phi_N$ hold.

$$\begin{aligned}\phi_1 \vee \phi_2 &::= \neg(\neg\phi_1 \wedge \neg\phi_2)\\ \phi_1 \Rightarrow \phi_2 &::= \neg(\phi_1 \wedge \neg\phi_2)\\ \phi_1 \equiv \phi_2 &::= (\phi_1 \Rightarrow \phi_2) \wedge (\phi_2 \Rightarrow \phi_1)\\ \phi_1 \not\equiv \phi_2 &::= \neg(\phi_1 \equiv \phi_2)\\ \mathrm{MPS}(\phi) &::= \mathrm{MCS}(\neg\phi)\\ \mathrm{SUP}(e) &::= \mathrm{IDP}(e, e_{top})\end{aligned}$$

$$\underset{\bowtie k}{\mathrm{Vot}}(\phi_1, \ldots, \phi_N) ::= \bigvee_{\substack{U \subseteq \{1,\ldots,N\}\\ |U| \bowtie k}} \left( \bigwedge_{u \in U} \phi_u \right) \wedge \left( \bigwedge_{u \in \{1,\ldots,N\} \setminus U} \neg\phi_u \right)$$

Here, we have $k \leq N$ and $\bowtie \in \{<, \leq, =, \geq, >\}$.

### *B. Semantics*

The semantics of *BFL* is given by the satisfaction relation $\models$, which expresses whether a formula $\phi$ holds for a FT $T$ under the status vector $\overline{b} = (b_1, \ldots, b_k)$. The satisfaction for events $e$, Boolean connectives and setting evidence are straightforward. For the MCS operator, we have $\overline{b}, T \models \mathrm{MCS}(\phi)$ iff $\overline{b}$ is a minimal vector that satisfies $\phi$, i.e., $\overline{b}$ satisfies $\phi$, but if we change some $b_i = 0$ into $b_i = 1$, then the vector no longer satisfies $\phi$. In particular, $\overline{b}, T \models \mathrm{MCS}(e_{top})$ means that $\overline{b}$ is a minimal cut set for the TLE of $T$.

$$\begin{aligned}
&\overline{b}, T \models e && \text{iff } \Phi_T(\overline{b}, e) = 1\\
&\overline{b}, T \models \neg\phi && \text{iff } \overline{b}, T \not\models \phi\\
&\overline{b}, T \models \phi \wedge \phi' && \text{iff } \overline{b}, T \models \phi \text{ and } \overline{b}, T \models \phi'\\
&\overline{b}, T \models \phi[e_i \mapsto 0] && \text{iff } \overline{b'}, T \models \phi \text{ with } \overline{b'} = (b'_1, \ldots, b'_n) \text{ where}\\
& && \quad b'_i = 0 \text{ and for } j \neq i \text{ we have } b'_j = b_j\\
&\overline{b}, T \models \phi[e_i \mapsto 1] && \text{iff } \overline{b'}, T \models \phi \text{ with } \overline{b'} = (b'_1, \ldots, b'_n) \text{ where}\\
& && \quad b'_i = 1 \text{ and } b'_j = b_j \text{ for } j \neq i\\
&\overline{b}, T \models \mathrm{MCS}(\phi) && \text{iff } \overline{b}, T \models \phi \wedge (\neg\exists \overline{b'}.\ \overline{b'} \subset \overline{b} \wedge \overline{b'}, T \models \phi)
\end{aligned}$$

With $[\![\phi]\!]$ we denote the *satisfaction set* of vectors for $\phi$, i.e., the set containing all status vectors that satisfy $\phi$. We define semantics for the second syntactic layer as:

$$\begin{aligned}
&T \models \exists\phi && \text{iff } \exists\overline{b}.\ \overline{b}, T \models \phi\\
&T \models \forall\phi && \text{iff } \forall\overline{b}.\ \overline{b}, T \models \phi\\
&T \models \mathrm{IDP}(\phi, \phi') && \text{iff } \mathrm{IBE}(\phi) \cap \mathrm{IBE}(\phi') = \emptyset
\end{aligned}$$

Here IBE is the set of influencing basic events defined as:

$$\begin{aligned}\mathrm{IBE}(\phi) = \{&e \in \mathrm{BE} \mid \exists\overline{b}.\ \overline{b}, T \models \phi[e \mapsto 0] \text{ and}\\ &\overline{b}, T \not\models \phi[e \mapsto 1] \text{ or vice-versa}\}\end{aligned}$$

Thus, $\mathrm{IBE}(\phi)$ returns the set of BEs that influence the truth value of $\phi$. Two events are independent if they do not share any IBEs.

**Example 1.** If we wanted to know whether the failure of the AND-gate *CP* in Fig. 1 implies the failure of the TLE *CP*/*R* for every possible truth assignment on BEs we would write $\forall(CP \Rightarrow CP/R)$. If we wanted to check whether it exists an assignment such that both *CP* and *CR* fail, we would write $\exists(CP \wedge CR)$. The same holds for $\vee$. Finally, if we wanted to check whether a specific vector $\overline{b}$ represents a MCS/MPS for *CP*/*R*, given $\overline{b}$ we would write $\mathrm{MCS}(CP/R)/\mathrm{MPS}(CP/R)$.

## IV. COVID Case Study: modeling

To illustrate our logic, we formulate several properties for a FT that models COVID-19 infection risks at construction sides. This FT is a slightly modified version of [26], and is displayed in Fig. 2.

**COVID-19 fault tree:** The TLE for this tree represents a COVID-19 infected worker on site, abbreviated *IWoS*. As shown in Fig. 2, the FT considers events in several categories: COVID-19 pathogens and reservoirs (i.e., germs and objects carrying the virus); their mode of transmissions; the presence of susceptible hosts, infected objects and workers; physical contacts; the status, as well as human errors. Note that Figure 2 contains several repeated basic events: *IT*, *PP*, *H1* and *IW* occur at multiple places in the tree and refer to the same basic event. This TLE *IWoS* is refined via an AND-gate with three children.

Thus, for the TLE to occur the following must happen: COVID pathogens/COVID infected objects must exist, there has to be a susceptible host and COVID pathogens must be transmitted in some way to this host. These three events are captured by corresponding subtrees: the purple OR-gate $CP/R$ refines the existence of COVID pathogens/COVID infected objects, the OR-gate $MoT$ in teal refines modes of transmission and the AND-gate $SH$ in orange refines the presence of a susceptible host. In order to further refine these trees in a realistic way, the FT contains duplicated leaves, e.g., *Physical Proximity* ($PP$) occurs twice.

**Properties:** Following, we specify some properties using natural language and present the corresponding *BFL* formulae:

- Is an infected surface sufficient for the transmission of COVID?

$$\forall(\mathit{IS} \Rightarrow \mathit{MoT})$$

- Does the occurrence of Mode of Transmission require human errors?

$$\forall(\mathit{MoT} \Rightarrow (\mathit{H1} \vee \mathit{H2} \vee \mathit{H3} \vee \mathit{H4} \vee \mathit{H5}))$$

- Is an object disinfection error sufficient for the occurrence of TLE?

$$\forall(\mathit{H4} \Rightarrow \mathit{IWoS})$$

- Are at least 2 human errors sufficient for the occurrence of TLE?

$$\forall(\underset{\geq 2}{\mathrm{Vot}}(\mathit{H1}, \ldots, \mathit{H5}) \Rightarrow \mathit{IWoS})$$

- What are all the MCSs for TLE that include errors in disinfecting objects?

$$[\![\mathrm{MCS}(\mathit{IWoS}) \wedge \mathit{H4}]\!]$$

- Is not committing any human error sufficient to prevent the occurrence of TLE?

$$\exists \mathrm{MPS}(\mathit{IWoS})[\underset{e_i \in \mathtt{BE} \setminus \{\mathit{H1}, \ldots, \mathit{H5}\}}{\mathit{H1} \mapsto 0, \mathit{H2} \mapsto 0, \mathit{H3} \mapsto 0,} \\ \mathit{H4} \mapsto 0, \mathit{H5} \mapsto 0, e_i \mapsto 1]$$

- What are the minimal ways to prevent the TLE?

$$[\![\mathrm{MPS}(\mathit{IWoS})]\!]$$

- Are a contact with an infected object and a contact with an infected surface independent scenarios?

$$\mathrm{IDP}(\mathit{CIO}, \mathit{CIS})$$

- Is physical proximity superfluous for the occurrence of TLE?

$$\mathrm{SUP}(\mathit{PP})$$

A thorough analysis of these properties is conducted in Sec. VII.

## V. Model Checking Algorithms

We present two model checking procedures for *BFL*.

1) Given a specific vector $\overline{b}$, a FT $T$ and a formula $\chi$, check if $\overline{b}, T \models \chi$ (Sec. V-C),
2) Given a FT $T$ and a formula $\chi$, compute all vectors $\overline{b}$ such that $\overline{b}, T \models \chi$ (Sec. V-D),

Furthermore, we offer some observations on the following problem: given a Boolean vector $\overline{b}$ and a formula $\chi$, compute a FT $T$ such that $\overline{b}, T \models \chi$ (Sec. V-E). Note that the first question, checking if $\overline{b}, T \models \chi$ holds, is trivial if $\chi = \phi$ is a level-1 formula that does not contain an MCS or MPS operator. In that case, we can simply substitute the values of the $\overline{b}$ in $\phi$ and see if the Boolean expression evaluates to true. For the other cases, the computation becomes more complex, and we resort to binary decision diagrams (BDDs). Concretely, to translate formulae to BDDs, we identify FT elements that appear in a given formula. We selectively construct BDDs for these elements and store the resulting BDDs (see Algo. 1): by doing so, we reduce computation time in case they are needed for other formulae. Finally, we manipulate these BDDs to reflect semantics of the operators in *BFL*. This translation to BDDs constitutes a formal ground that permits to address the aforementioned procedures in a uniform way. In particular, once the BDD for the formula is obtained, we can address the first and second scenarios. In the former, we walk down the BDD from the root node following truth assignments given in a specific vector $\overline{b}$. In the latter — where no vector is given — we collect every path that leads to the terminal `1` to compute all satisfying vectors $[\![\overline{b}]\!]$ for the given formula.

### A. Binary decision diagrams

BDDs are directed acyclic graphs (DAGs) that offer a compact way to represent Boolean functions [27] by reducing redundancy. BDD's size can grow linearly in the number of variables and at worst exponentially, depending on variable's ordering. In practice, BDDs are heavily used in various application areas, including in FT analysis and in their security-related counterpart, attack trees (ATs) [5, 28].

Formally, a BDD is a rooted DAG $\mathbf{B}_f$ that represents a Boolean function $f \colon \mathbb{B}^n \to \mathbb{B}$ over variables $\mathit{Vars} = \{x_i\}_{i=1}^{n}$. The outcomes of $f$ - `0` or `1` - are represented by the terminal nodes of $\mathbf{B}_f$. A non terminal node $w \in W$ represents a subfunction $f_w$ of $f$ via its Shannon expansion [27]. In other words, $w$ is equipped with a variable $\mathit{Lab}(w) \in \mathit{Vars}$ and two children: $\mathit{Low}(w) \in W$, that represents $f_w$ in case that the variable $\mathit{Lab}(w)$ is set to `0`; and $\mathit{High}(w)$, representing $f_w$ if $\mathit{Lab}(w)$ is set to `1`.

**Definition 5.** A *BDD* is a tuple $\mathbf{B} = (W, \mathit{Low}, \mathit{High}, \mathit{Lab})$ over a set $\mathit{Vars}$ where:

- The *set of nodes* $W$ is partitioned into terminal nodes ($W_t$) and non terminal nodes ($W_n$);
- $\mathit{Low} \colon W_n \to W$ maps each node to its *low child*;

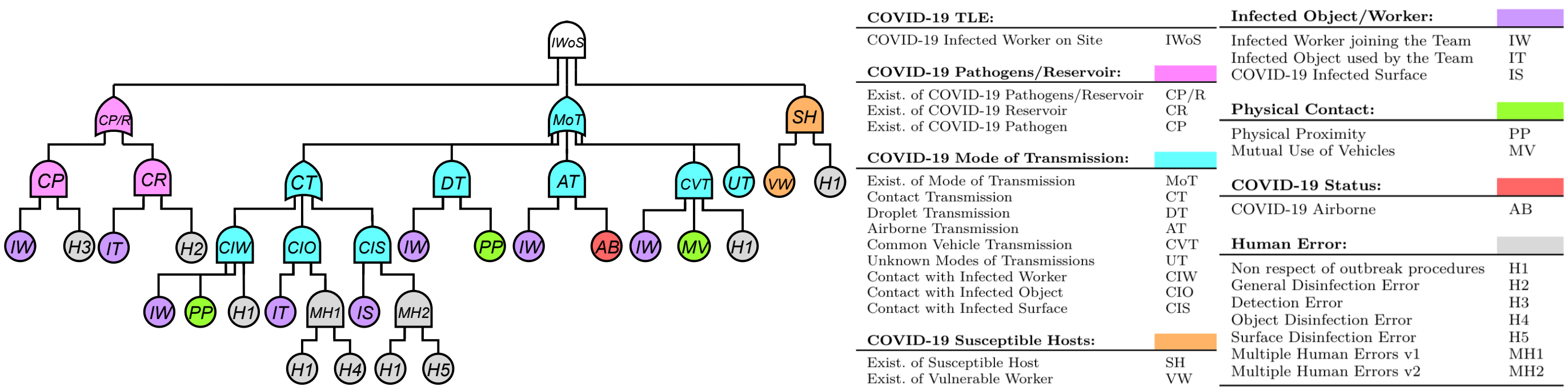

Figure 2: COVID-19 FT.

- *High*: $W_n \to W$ maps each node to its *high child*;
- *Lab*: $W \to \{0, 1\} \cup Vars$ maps terminal nodes to Booleans, and non terminal nodes to variables:
$$Lab(w) \in \begin{cases} \{0,1\} & \text{if } w \in W_t, \\ Vars & \text{if } w \in W_n. \end{cases}$$

Moreover, $\mathbf{B}$ satisfies the following constraints:

- $(W, E)$ is a connected DAG, where
$$E = \{(w, w') \in W^2 \mid w' \in Low(w) \cup High(w)\};$$
- $\mathbf{B}$ has a unique root, denoted $R_B$:
$$\exists!\, R_B \in W.\ \forall w \in W_n.\ R_B \notin Low(w) \cup High(w).$$

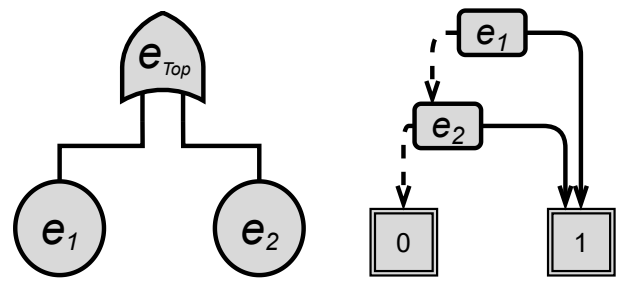

Figure 3: A simple FT (OR-gate) and its BDD.

**Reduced ordered BDDs:** Conventionally, the term BDDs is often used as a shorthand acronym for reduced ordered BDDs, or ROBDDs [29, 30]. ROBDDs are BDDs in which the variables occur in a given order on all the paths of the BDD, no two distinct nodes have the same variable name and (high and low) successors, and no single node has identical (high and low) successors [27]. As such, a total order $<$ over the variables is needed. Formally, this means that:

- *Vars* are equipped with a total order, $\mathbf{B}_f$ is thus defined over a pair $\langle Vars, < \rangle$;
- the variable of a node is of lower order than its children: $\forall w \in W_n.\ Lab(w) < Lab(Low(w)), Lab(High(w))$;
- the children of non terminal nodes are distinct nodes;
- all terminal nodes are distinctly labelled.

ROBDDs exhibit the following characteristics: 1) there are exactly two terminal nodes: $W_t = \{\bot, \top\}$, with $Lab(\bot) = 0$ and $Lab(\top) = 1$; 2) the label of the root node $R_B$ has the lowest order; 3) in any two paths from $R_B$ to $\bot$ or $\top$, variables appear in the same (increasing) order.

*B. Algorithm 1: Translating FTs/formulae to BDDs*

**Translations:** As a first step, we will define BDD translations for both FTs and formulae in the logic. As previously mentioned, to translate formulae to BDDs, we first identify FT elements that appear in a given formula. We selectively construct BDDs for these elements and store them. This allows us to reduce computation time in case they are needed to construct BDDs for other formulae that include the same FT elements. Finally, we manipulate these BDDs to reflect semantics of the operators in *BFL*. Operations between BDDs are represented by **bold** operands e.g., $\boldsymbol{\wedge}, \boldsymbol{\vee}$. Operands can be applied by following Algo. 5.15 [29] Apply and subsequently applying Algo. 5.3 Reduce [29] to ensure the resulting BDD is reduced. Following, we assume to always apply Reduce after Apply. Quantification can be achieved by applying Algo. 5.15 [29] Apply and Algo. 5.20 [29] Restrict as shown in Theorem 5.23 [29]. In particular, given a set of variables $\mathtt{V} = \{v_1, \ldots, v_n\}$, existential quantification can be defined as follows:

$$\begin{aligned} \exists v.\mathbf{B} &= \textsc{Restrict}(\mathbf{B}, v, 0) \boldsymbol{\vee} \textsc{Restrict}(\mathbf{B}, v, 1) \\ \exists V.\mathbf{B} &= \exists v_1.\exists v_2. \ldots \exists v_n.\mathbf{B} \end{aligned}$$

**Translating FTs to BDDs:** Firstly, we will define a translation from FTs to BDDs. In the following paragraphs we assume $Vars = \mathtt{V} \dot{\cup} \mathtt{V}'$, where the set of variables $\mathtt{V} = \mathtt{BE}$ and the set of primed variables $\mathtt{V}' = \{e' | e \in \mathtt{BE}\}$. Furthermore, we define $Var_{\mathbf{B}}\colon \mathtt{BDD} \to Vars$ to be a function that returns variables occurring in a BDD. Let $\Psi_{FT}\colon \mathtt{E} \to \mathtt{BDD}$ be a function that takes elements of a FT as input and maps them to BDDs. $\Psi_{FT}$ is defined as follows:

**Definition 6.** The translation function of a FT $T$ is a function $\Psi_{FT_T}\colon \mathtt{E} \to \mathtt{BDD}$ that takes as input an element $e \in \mathtt{E}$. With $e' \in ch(e)$, we can define $\Psi_{FT_T}$:

$$\Psi_{FT_T}(e) = \begin{cases} \overline{\mathbf{B}}(e) & \text{if } e \in \mathtt{BE} \\ \bigvee \Psi_{FT_T}(e') & \text{if } e \in \mathtt{IE} \text{ and } t(e) = \mathtt{OR} \\ \bigwedge \Psi_{FT_T}(e') & \text{if } e \in \mathtt{IE} \text{ and } t(e) = \mathtt{AND} \\ \displaystyle\bigvee_{\substack{n_1,\ldots,n_k \\ n_1 < \ldots < n_k}} \bigwedge_{i=1}^{k} \Psi_{FT_T}(e'_{n_i}) & \text{if } e \in \mathtt{IE} \text{ and } t(e) = \mathtt{VOT}(k/N) \end{cases}$$

Where $\overline{\mathbf{B}}(v)$ is a BDD with a single node in which $Low(v) = 0$ and $High(v) = 1$.

Fig. 3 represents a simple FT i.e., a single OR-gate, and its translation to BDDs.

**Translating formulae:** Knowing how to compute BDDs for FTs, we can now show how to manipulate them in order to mirror *BFL* operators. I.e., given $\Psi_{FT}$ and a FT $T$, for every *BFL* formula $\chi$ in the set of *BFL* formulae $X$ we can now show a translation to BDDs $\underline{\mathbf{B}}_T \colon X \to \mathtt{BDD}$ in Algo. 1. The implementation of this procedure would abide the dynamic programming standards: by simple caching, we would be able to reuse the translation of (sub)trees and (sub)formulae between different analyses without recomputing them each time anew.

---

**Algorithm 1** Given $\chi$ and $T$, compute $\underline{\mathbf{B}}_T(\chi)$

---

**Input:** FT $T$, formula $\chi$

**Output:** $\underline{\mathbf{B}}_T(\chi)$

**Method:** Compute $\underline{\mathbf{B}}_T(\chi)$ according to the recursion scheme below. Store intermediate results $\underline{\mathbf{B}}_T(\cdots)$ and $\Psi_{FT_T}(\cdots)$ in a cache in case they are used several times.

**Recursion scheme:**

$\underline{\mathbf{B}}_T(e)$ : $\Psi_{FT_T}(e)$

$\underline{\mathbf{B}}_T(\neg\phi)$ : $\neg(\underline{\mathbf{B}}_T(\phi))$

$\underline{\mathbf{B}}_T(\phi \wedge \phi')$ : $\underline{\mathbf{B}}_T(\phi) \wedge \underline{\mathbf{B}}_T(\phi')$

$\underline{\mathbf{B}}_T(\phi[e_i \mapsto 0])$ : $\textsc{Restrict}(\underline{\mathbf{B}}_T(\phi), e_i, 0)$

$\underline{\mathbf{B}}_T(\phi[e_i \mapsto 1])$ : $\textsc{Restrict}(\underline{\mathbf{B}}_T(\phi), e_i, 1)$

$\underline{\mathbf{B}}_T(\mathrm{MCS}(\phi))$ : $\underline{\mathbf{B}}_T(\phi) \wedge (\neg\exists \mathtt{V}'.\mathtt{V}' \subset \mathtt{V} \wedge \underline{\mathbf{B}}_T(\phi)[\mathtt{V} \curvearrowright \mathtt{V}'])$ where:

$$\mathtt{V}' \subset \mathtt{V} \equiv \left(\bigwedge_k v'_k \Rightarrow v_k\right) \wedge \left(\bigvee_k v'_k \neq v_k\right)$$

$\underline{\mathbf{B}}_T(\exists\phi)$ : $\exists \mathtt{V}.\underline{\mathbf{B}}_T(\phi)$

$\underline{\mathbf{B}}_T(\forall\phi)$ : $\neg\exists \mathtt{V}.\neg\underline{\mathbf{B}}_T(\phi)$

$\underline{\mathbf{B}}_T(\mathrm{IDP}(\phi, \phi'))$: $\mathtt{1}$ iff $\{Var_{\mathbf{B}}(\underline{\mathbf{B}}_T(\phi)) \cap Var_{\mathbf{B}}(\underline{\mathbf{B}}_T(\phi'))\} = \emptyset$

---

where $\underline{\mathbf{B}}_T(\phi)[\mathtt{V} \curvearrowright \mathtt{V}']$ indicates the BDD $\underline{\mathbf{B}}_T(\phi)$ in which every variable $v_k \in \mathtt{V}$ is renamed to its primed $v'_k \in \mathtt{V}'$. To perform existential and universal quantification over formulae in this translation - i.e., utilizing the second layer in the syntax of *BFL* - it would also suffice to check if the resulting BDD for the formula $\chi$ is not equivalent to the terminal node $\mathtt{0}$ in the case of existential quantification and to check if the BDD for $\chi$ is equivalent to the BDD for the terminal node $\mathtt{1}$ in case of universal quantification.

### *C. Algorithm 2: Model checking BFL over a FT and a $\overline{b}$*

**Overview:** As mentioned, given a specific vector $\overline{b}$, a FT $T$ and a formula $\chi$, we want to check if $\overline{b}, T \models \chi$. To do so, we translate the given formula to a BDD and then we walk down the BDD from the root node following truth assignments given in the specific vector $\overline{b}$.

---

**Algorithm 2** Check if $\overline{b}, T \models \chi$, given $\overline{b}$, $T$ and $\chi$.

---

**Input**: Boolean vector $\overline{b}$ , FT $T$ and a formula $\chi$

**Output:** *True* iff $\overline{b}, T \models \chi$, *False* otherwise

**Method:**

**compute** $\underline{\mathbf{B}}_T(\chi)$ via Algo. 1

Starting from BDD root,

**while** current node $w_i$ of $\underline{\mathbf{B}}_T(\chi) \notin W_t$ **do**:

  **if** $b_i \in \overline{b} = 0$ **then**:

    $w_i = Low(w_i)$

  **else if** $b_i \in \overline{b} = 1$ **then**:

    $w_i = High(w_i)$

  **end if**

**end while**

**if** $Lab(w_i) = \mathtt{0}$ **then**:

  **return** False

**else if** $Lab(w_i) = \mathtt{1}$ **then**:

  **return** True

**end if**

---

**Algo. 2:** Algo. 2 shows an algorithm to check whether $\overline{b}, T \models \chi$, given a status vector $\overline{b}$, a FT $T$ and a formula $\chi$. A BDD for the formula $\chi$ is computed with regard to the structure function of the given FT $T$ i.e., we compute $\underline{\mathbf{B}}_T(\chi)$ as per Algo. 1. Subsequently, the algorithm walks down the BDD following the Boolean assignments given in $\overline{b}$: if the i-th element of $\overline{b}$ is set to 0 then the next node in the path will be given by $Low(w_i)$, if it is set to 1 then the next node will be $High(w_i)$. When the algorithm reaches a terminal node it returns *True* if its value is one - i.e., if $\overline{b}, T \models \chi$ - and *False* otherwise.

**Example 2.** To showcase this procedure, we present the smallest possible example that still retains significance, by choosing a FT $T$ with a single OR-gate as a TLE with two BEs as children. We will assume that our formula $\chi$ is $\mathrm{MCS}(e_{top})$ and that the given vector $\overline{b}$ is equal to $(0, 1)$ for $b_1$ and $b_2$ respectively. Fig. 3 shows a representation of this simple FT next to its BDD. We then compute the BDD $\underline{\mathbf{B}}_T(\mathrm{MCS}(e_{top}))$ for our formula and walk down the graph following the low edge for $w_1$ and the high edge for $w_2$, as indicated by the values in $\overline{b}$. We end the algorithm in the terminal node $\mathtt{1}$, thus knowing that $\overline{b}, T \models \mathrm{MCS}(e_{top})$. The figure in this paragraph represents the BDD for $MCS(e_{top})$ for the given FT: the path representing vector $\overline{b}$ is highlighted.

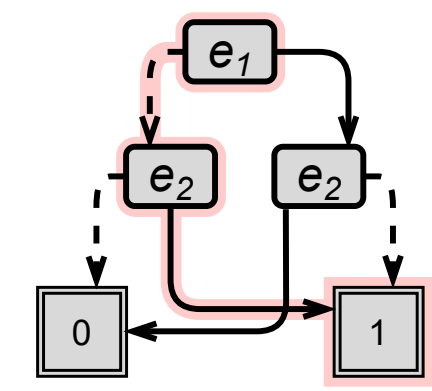

### *D. Algorithm 3: Computing all satisfying vectors*

**Overview:** In this subsection, we will address a different scenario: given a FT $T$ and a formula $\chi$, we want to compute all vectors $\overline{b}$ such that $\overline{b}, T \models \chi$. In this scenario no vector is given. Thus, we need to construct the BDD for the given formula and then collect every path that leads to the terminal $\mathtt{1}$ to compute all satisfying vectors $[\![\overline{b}]\!]$ for the given formula.

**Algo. 3:** To achieve the desired outcome we will construct the BDD $\underline{\mathbf{B}}_T(\chi)$ for the given formula following Algo. 1. Then, the algorithm will walk down the BDD and store all the paths that lead to the terminal node 1. These paths represent all the status vectors that satisfy our formula $\chi$. The value for the elements of each vector is set to 0 or 1 if the stored path follows respectively the low or high edge of the collected elements of the BDD. After computing the BDD for a given $\chi$, AllSat [27] will achieve the desider outcome. This algorithm returns exactly all the satisfying assignments for a given BDD, i.e., in our case, all the vectors that satisfy our formula.

**Example 3.** To showcase this procedure, let us present an example. We consider the same FT presented in Ex. 2, for which we do not have a specific vector $\overline{b}$ to check against. We choose again MCS($e_{top}$) as our formula $\chi$. In order to compute all the MCSs for $e_{top}$, we first construct the corresponding BDD $\underline{\mathbf{B}}_T$(MCS($e_{top}$)), then we apply AllSat [29]. We walk down the BDD and collect all the nodes in paths that lead to the terminal labelled with 1. By doing so, we obtain two different vectors $\overline{b}$ and $\overline{b'}$ which respectively contain $(0,1)$ and $(1,0)$ as the values for $b_1, b_2$ and $b'_1, b'_2$, representing the different assignments that satisfy our formula given the specific FT $T$. The figure in this paragraph represents the BDD for MCS($e_{top}$): both the paths that represent satisfying assignments for $\chi$ are highlighted.

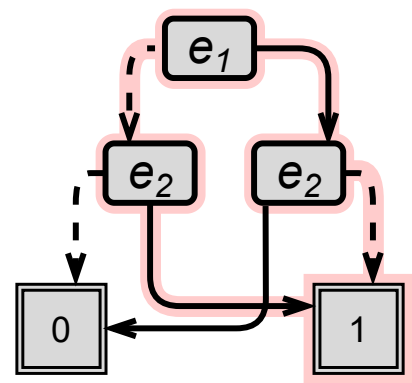

*E. Computing a satisfying fault tree*

**Overview:** In this section we will briefly discuss the following problem: given a vector $\overline{b}$ and a formula $\chi$, we want to compute at least one FT $T$ such that $\overline{b}, T \models \chi$.

**Observations:** A more trivial approach could consist in trying to satisfy the given formula by keeping the values of BEs fixed - as given in $\overline{b}$ - and trying out all possible truth assignments for the other variables in the given $\chi$, until $\chi$ is satisfied. However, this procedure does not guarantee that the resulting structure of the FT will be meaningful. More complex procedures - out of the scope of this paper - can infer the structure of a FT from given vector(s) $\overline{b}$ while also considering other relevant properties of the tree, like its complexity and the number of elements e.g., employing genetic algorithms [31].

## VI. Counterexamples

**Overview:** If a formula does not hold, it is important to know why this property fails. Counterexamples provide such diagnostic information. Given a formula $\chi$ that is not satisfied by the vector $\overline{b}$, a counterexample is simply a new vector $\overline{b'}$ with minimal modifications s.t. it does satisfy $\chi$. We present an algorithm to compute such counterexamples (see Algo. 4). In so doing, we can provide counterexamples for significant properties that can be captured by *BFL*. Since the most common qualitative analyses on FTs involve MCSs and MPSs, we will focus on counterexamples and present *patterns* for these operators. For example, suppose we believe that the failure of $\{IW, H3, IT\}$ is a MCS for the TLE in $T$ on the right. Clearly, $\{IW, H3, IT\}$ is not an MCS for $T$: it is a cut set but not minimal. As highlighted in the figure, a suitable counterexample is a MCS contained in $\{IW, H3, IT\}$, i.e., $\{IW, H3\}$.

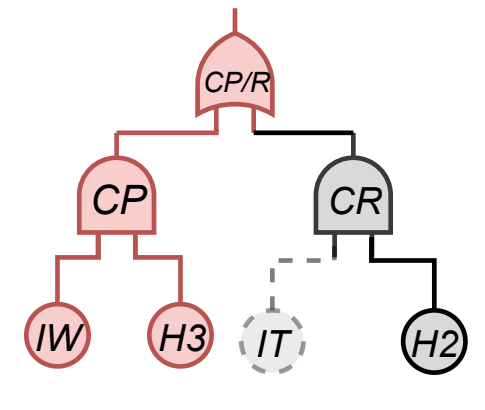

**Counterexamples:** We present our algorithm to compute counterexamples and a corresponding definition:

---

**Algorithm 4** Compute a counterexample $\overline{b'}$ s.t. $\overline{b'}, T \models \chi$, given $\overline{b}$, $T$ and $\chi$, s.t. $\overline{b}, T \not\models \chi$.

---

**Input**: Boolean vector $\overline{b}$, FT $T$ and a formula $\chi$
**Output:** counterexample vector $\overline{b'}$
**Method:**
**compute** $\underline{\mathbf{B}}_T(\chi)$ via Algo. 1
**if** $1 \notin W_t$ **then**:
  **return**
**end if**
Starting from BDD root,
**while** current node $w_i$ of $\underline{\mathbf{B}}_T(\chi) \notin W_t$ **do**:
  $w'_i = w_i$
  **if** $b_i \in \overline{b} = 0$ **then**:
    $w_i = Low(w_i)$
    $b'_i = 0$
  **else if** $b_i \in \overline{b} = 1$ **then**:
    $w_i = High(w_i)$
    $b'_i = 1$
  **end if**
  **if** $Lab(w_i) = 0$ **then**:
    **if** $b_i \in \overline{b} = 0$ **then**:
      $w_i = High(w'_i)$
      $b'_i = 1$
    **else if** $b_i \in \overline{b} = 1$ **then**:
      $w_i = Low(w'_i)$
      $b'_i = 0$
    **end if**
  **end if**
**end while**
set all values $b'_i$ which have not been set to the same values as according $b_i$
**return** $\overline{b'}$

---

**Definition 7.** Given a FT $T$, a Boolean vector $\overline{b}$ and a formula $\chi$ such that $\overline{b}, T \not\models \chi$, a *counterexample* is a new vector $\overline{b'}$ such that $\overline{b'}, T \models \chi$. Further, for all $b'_i$ with $b'_i \neq b_i$ we have that $(b'_1, \ldots, b'_{i-1}, b_i, b'_{i+1}, \ldots), T \not\models \chi$.

The reason for Def. 7 is as follows: If we have a vector $\overline{b}$ such that with this vector the FT and the formula are not fulfilled, we want modify $\overline{b}$ to a $\overline{b'}$ in such a way that the FT and formula is indeed fulfilled. We also do not want to consider a completely different vector $\overline{b'}$, but one in which the modifications performed are as small as possible. For this reason, we require that we cannot change the value of variables in $\overline{b'}$ such that they are the same as in $\overline{b}$ without invalidating the FT/formula combination. Algo. 4 works as follows: We compute $\underline{\mathbf{B}}_T(\chi)$ as per Algo. 1. Starting with the first variable in the variable order of the BDD, we walk down the BDD. We follow the *Low* branch if the according value in $\overline{b}$ is 0 and the *High* branch if it is 1. If we do not end up at the `0` node, we continue with the next variable. If however we end up at the `0` node, we revise our decision and instead take the other branch of the BDD node. Decisions taken when descending the BDD are stored in vector $\overline{b'}$. For all BDDs different from `0`, we thus compute a $\overline{b'}$ such that 1) it fulfils the FT/formula combination and 2) the variable values where it differs from $\overline{b}$ cannot be changed without invalidating the formula.

**Definition 8.** A *pattern* is a *BFL* formula where non-terminal symbols might be present. A pattern *matches* a formula whenever a valid *BFL* formula can be generated from that pattern.

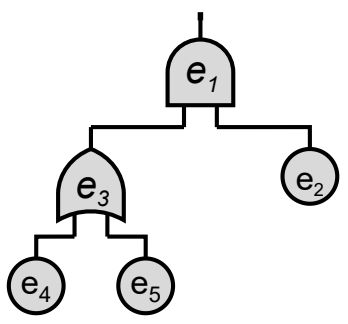

In light of Algo. 4 and making use of *patterns*, we present a vector $\overline{b}$, a tree $T$ — represented on the right — and one or more example formulae $\chi$. We then construct a representation of failure propagation of BEs in $T$ with respect to the truth assignments in $\overline{b}$. In case $\overline{b}$ does not satisfy $\chi$ for $T$, we present a new status vector that satisfies $\chi$ for $T$, via Algo. 4. Furthermore, we represent this counterexample and show how failure propagation on $T$ differs between the example and its corresponding counterexample. Table I collects example formulae, their visualization with respect to given vectors, and corresponding counterexamples, both visualized in $T$ and represented as vectors. Let us consider MCSs and MPSs. There are two general cases in which $\overline{b}$ does not satisfy MCS($\phi$) or MPS($\phi$):

1) If $\overline{b}$ is a cut set/path set for $\chi$, but is not minimal.
2) If $\overline{b}$ is not a MCS/MPS for $\chi$, i.e. it lacks elements.

**Patterns 1 and 2:** Let $pattern_1$ ::= MCS($\phi$) and $pattern_2$ ::= MPS($\phi$). Consider $T$ and let the vector $\overline{b}$ represent the ordered Boolean values for $e_2, e_4, e_5$. For $\chi$ = MCS($e_1$) and $\overline{b}$ representing respectively the failure of $e_4$ and $e_2, e_4, e_5$ we would have the following vectors as counterexamples: $\overline{b}' = (1,1,0)$ for the former and $\overline{b}' = (1,0,1)$ for the latter. While for $\chi = \text{MPS}(e_1)$ and $\overline{b}$ representing respectively events $e_4$ and $e_2, e_4, e_5$ being operational, we would have the following vectors as counterexamples: $\overline{b}' = (1,0,0)$ for the former and $\overline{b}' = (0,1,1)$ for the latter.

| Pattern | Example $\chi$ | Ex. vector | Ex. representation | Cex. vector | Cex. representation |
|---|---|---|---|---|---|
| $pattern_1 ::= \mathrm{MCS}(\phi)$ | $\mathrm{MCS}(e_1)$ | $\overline{b} = (0,1,0)$ | | $\overline{b'} = (1,1,0)$ | |
| | | $\overline{b} = (1,1,1)$ | | $\overline{b'} = (1,0,1)$ | |
| $pattern_2 ::= \mathrm{MPS}(\phi)$ | $\mathrm{MPS}(e_1)$ | $\overline{b} = (1,0,1)$ | | $\overline{b'} = (1,0,0)$ | |
| | | $\overline{b} = (0,0,0)$ | | $\overline{b'} = (0,1,1)$ | |
| $pattern_3 ::= \mathrm{MCS}(\phi_1) \wedge \ldots \wedge \mathrm{MCS}(\phi_n)$ | $\mathrm{MCS}(e_1) \wedge \mathrm{MCS}(e_3)$ | $\overline{b} = (0,1,0)$ | | $\overline{b'} = (1,1,0)$ | |
| $pattern_4 ::= \mathrm{MPS}(\phi_1) \wedge \ldots \wedge \mathrm{MPS}(\phi_n)$ | $\mathrm{MPS}(e_1) \wedge \mathrm{MPS}(e_3)$ | $\overline{b} = (1,0,1)$ | | $\overline{b'} = (1,0,0)$ | |

Table I: A collection of patterns, exemplifications through given $\chi$ and $\overline{b}$, and the respective counterexamples.

**Patterns 3 and 4:** Let $pattern_3 ::= \text{MCS}(\phi_1) \wedge \ldots \wedge \text{MCS}(\phi_n)$ and $pattern_4 ::= \text{MPS}(\phi_1) \wedge \ldots \wedge \text{MPS}(\phi_n)$. Given a FT $T$ and a vector $\overline{b}$, these two patterns allow us to check whether a given $\overline{b}$ represents a MCS/MPS for $n$ gates in $T$. Given $T$ and $\overline{b}$ representing the failure of $e_4$, we can check if $\overline{b}$ is a MCS for both $e_1$ and $e_3$ i.e., $\chi = \text{MCS}(e_1) \wedge \text{MCS}(e_3)$. The failure of $e_4$ is not sufficient, thus we present a counterexample vector $\overline{b}' = (1,1,0)$. The same procedure applied to MPSs can be seen in $pattern_4$. Given the same FT and $\overline{b}$ representing $e_4$ being operational, we observe that $\overline{b}$ does not represent a MPS for both $e_1$ and $e_3$ i.e., $\chi = \text{MPS}(e_1) \wedge \text{MPS}(e_3)$. We present a counterexample $\overline{b}' = (1,0,0)$. Table I collects these examples, alongside their graphical representations.

## VII. COVID Case Study: Analysis

**Overview:** In this section we discuss the case study of $T$ in Fig. 2 in light of the procedures described in Sec. V and Sec. VI. For every property presented in Sec. IV, we discuss whether it holds for the given tree in Fig. 2. Furthermore, we present possible additional steps in the analysis of $T$ by leveraging the expressiveness of *BFL*. We then present a meaningful excerpt of Fig. 2. Storm model checker [32] is used in order to retrieve MCSs/MPSs whenever the corresponding operators appear in the following properties.

**Analysis:**

- Is an infected surface sufficient for the transmission of COVID? $\forall(IS \Rightarrow MoT)$

This property does not hold for the (sub)tree we are considering (see figure on the right). In fact, the failure of $IS$ alone is not sufficient for $MoT$ to fail. We can gather more information about the role $IS$ has in the failure of $MoT$ by computing MCSs for $MoT$ and filter for $IS$: $[\![\text{MCS}(MoT) \wedge IS]\!]$. The result is a single MCS containing $\{IS, H1, H5\}$. From this information we can infer that an infected surface paired with a surface disinfection error and the non-respect of outbreak procedures is sufficient for the transmission of COVID.

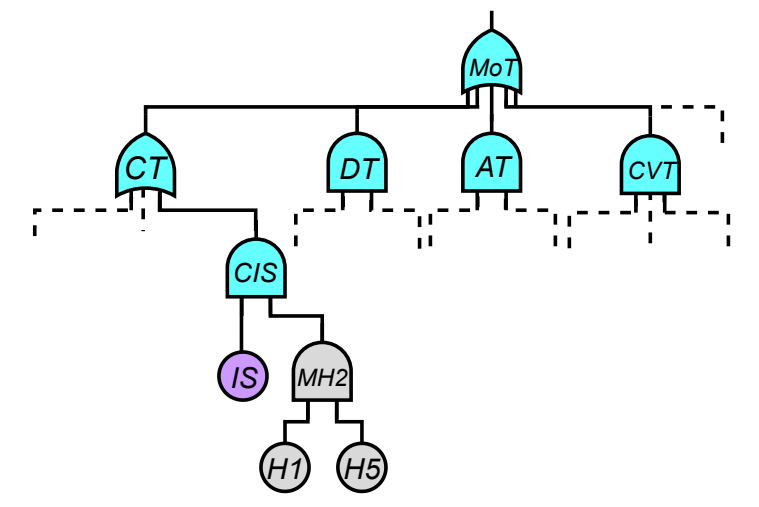

- Does the occurrence of Mode of Transmission require human errors? $\forall(MoT \Rightarrow (H1 \vee H2 \vee H3 \vee H4 \vee H5))$

Checking this property would return *False*. In fact, there are cases in which $MoT$ would occur without the need of any human error. E.g., in case of *Droplet Transmission* ($DT$) or *Airborne Transmission* ($AT$).

- Is an object disinfection error sufficient for the occurrence of TLE? $\forall(H4 \Rightarrow IWoS)$

This property does not hold for $T$. In fact, there are no cases in which the failure of $H4$ alone is sufficient for the failure of the TLE (see Fig. 4).

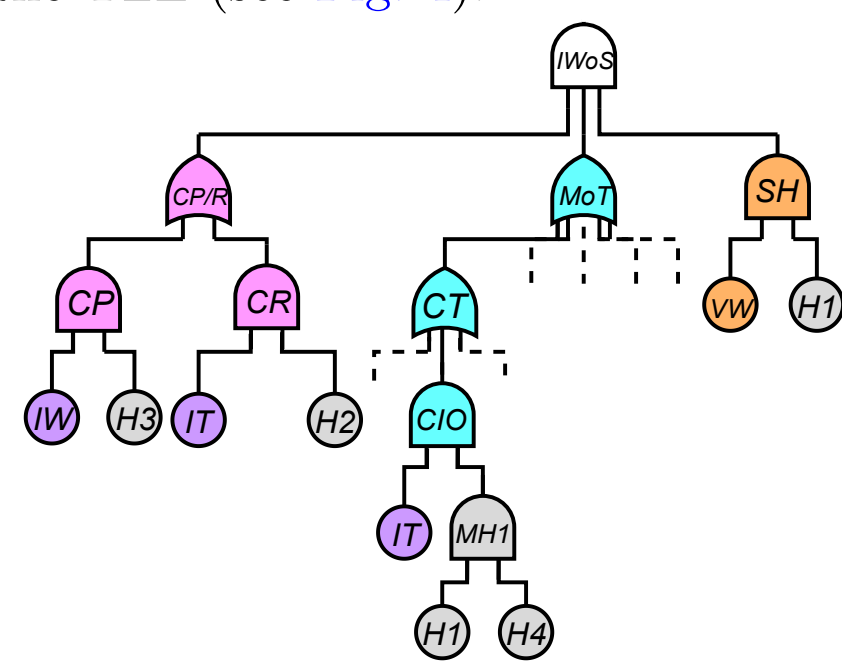

Figure 4: (Sub)tree for property 3 (excerpt from Fig. 2).

- Are at least 2 human errors sufficient for the occurrence of TLE? $\forall(\text{Vot}_{\geq 2}(H1, \ldots, H5) \Rightarrow IWoS)$

Two or more human errors are not sufficient for the occurrence of $IWoS$ (see Fig. 5). However, we can further investigate the role of human errors in the occurrence of TLE by asking which MCSs require the presence of human errors: $[\![(\text{MCS}(IWoS) \wedge H1) \vee \ldots \vee (\text{MCS}(IWoS) \wedge H5)]\!]$. By doing so we would obtain twelve MCSs. In spite of assessing the importance of human errors in the failure of TLE, the resulting number of MCSs can be slightly overwhelming to perform an accurate evaluation on. Thus, we might focus our assessment on single human errors. Our scope can be refined by the following property:

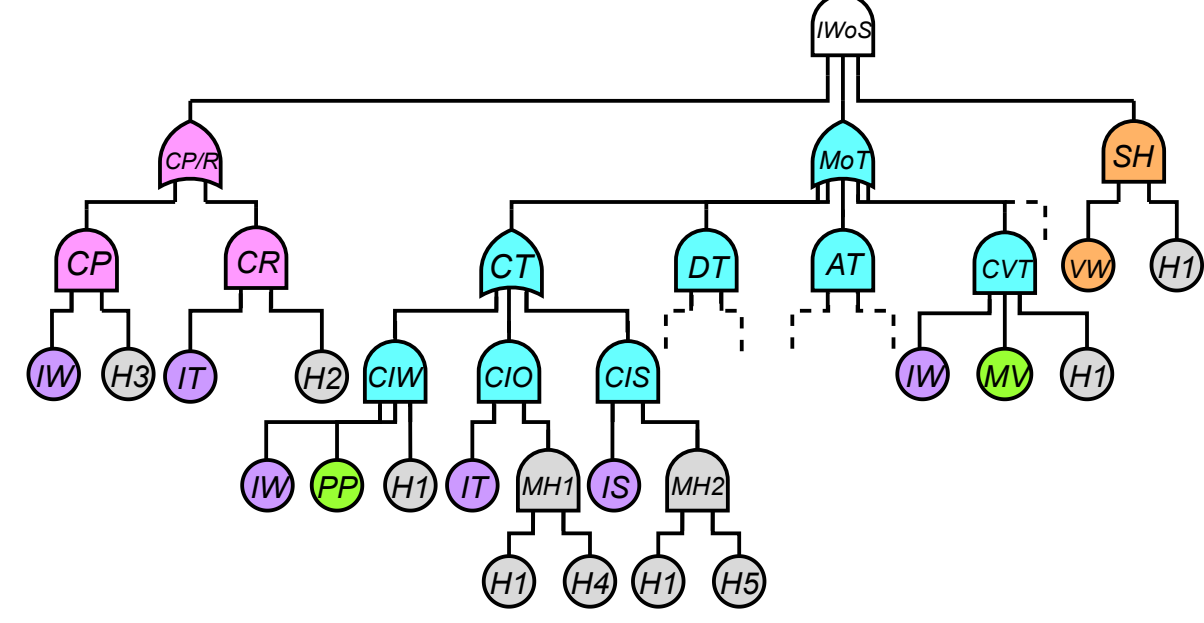

Figure 5: Tree for property 4 (excerpt from Fig. 2).

- What are all the MCSs for TLE that include errors in disinfecting objects? $[\![\text{MCS}(IWoS) \wedge H4]\!]$

Choosing a single error — e.g., $H4$ — allows us to narrow our objective and better isolate failure causes. By this query, we obtain all the MCSs that include $H4$, i.e., all the ways in which an infected worker can join the workplace (TLE) due to a series of actions containing a disinfection error (see Fig. 4): $\{IW, H3, IT, H1, H4, VW\}$ and $\{IT, H2, H1, H4, VW\}$. This analysis uncovered a potentially worrying MCS, heavy on human errors and with only two other elements needed to trigger a system-level failure, i.e., the *existence of a vulnerable worker* ($VW$) and an *infected object used by the team* ($IT$):

$\{IT, H2, H1, H4, VW\}$. Knowing this can prompt modifications in the workplace to mitigate the uncovered risk.

- Is not committing any human error sufficient to prevent the occurrence of TLE?

$$\exists \mathrm{MPS}(IWoS)[H1 \mapsto 0, H2 \mapsto 0, H3 \mapsto 0, \underset{e_i \in \mathtt{BE}\setminus\{H1,\dots,H5\}}{H4 \mapsto 0, H5 \mapsto 0, e_i \mapsto 1}]$$

The path set we are trying to construct in the formula above is not minimal, i.e., there are ways to prevent the occurrence of TLE by only addressing human errors (see Fig. 5), but including all five of them is not a minimal way to do so. In this setting, a counterexample would be constructed following $pattern_2$. In this case, the two vectors represented by the following MPSs are constructed: $\{H1\}$ and $\{H2, H3\}$. This means that if we can guarantee that either *H1* is not committed or *H2* alongside *H3* are not committed we will be sure that *IWoS* will not fail, i.e., an infected worker joining the team can be prevented either by respecting the outbreak procedures or by operational disinfection and detection of COVID on the workplace.

- What are all the minimal ways to prevent the occurrence of TLE? $[\![\mathrm{MPS}(IWoS)]\!]$

We can further extend our search on minimal ways to prevent the TLE to all the MPSs. This query returns all the MPSs for *IWoS*: $\{IW, IT\}$, $\{IW, H2\}$, $\{IW, H4, IS, UT\}$, $\{IW, H4, H5, UT\}$, $\{H3, IT\}$, $\{H3, H2\}$, $\{IT, PP, IS, AB, MV, UT\}$, $\{IT, PP, H5, AB, MV, UT\}$, $\{PP, H4, IS, AB, MV, UT\}$, $\{PP, H4, H5, AB, MV, UT\}$, $\{H1\}$ and $\{VW\}$. In doing so, we uncover interesting additional ways to prevent the occurrence of TLE, e.g., the MPS with a single element $\{VW\}$ or the ones with only two BEs $\{IW, H2\}$, $\{IW, IT\}$ and $\{H3, IT\}$. Some of these MPSs are of particular interest since they do not depend on human errors. This could further inform risk mitigation decisions, e.g., trying to reduce the risk of a vulnerable worker joining the team *VW*.

- Are a contact with an infected object and a contact with an infected surface always independent scenarios? $\mathrm{IDP}(CIO, CIS)$

As a last step, we might want to assess some events in our tree for their independence. E.g., considering the dependency between coming in contact with an infected object and with an infected surface (see excerpt on the right). By applying IDP we discover that *CIO* and *CIS* are not independent. In fact, if we return the influencing basic events IBEs for the two elements it becomes clear that their truth values both depend on *H1*, i.e., on the respect of outbreak procedures.

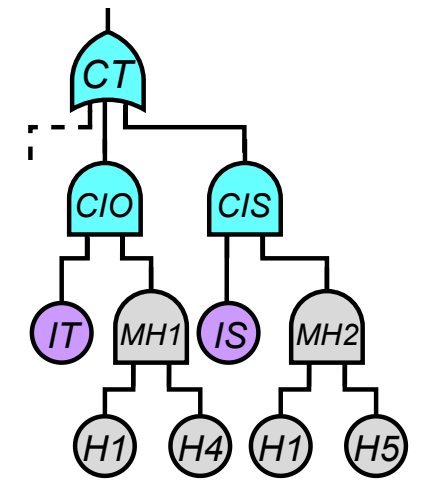

- Is physical proximity superfluous for the occurrence of TLE? $\mathrm{SUP}(PP)$

Finally, we can check whether physical proximity is superfluous for TLE (see Fig. 6). *PP* and *IWoS* are not independent: in fact, there are cases in which the failure of *IWoS* depends on the value of *PP*. Thus, *PP* is not a superfluous element in *T* and must not be removed from its leaves.

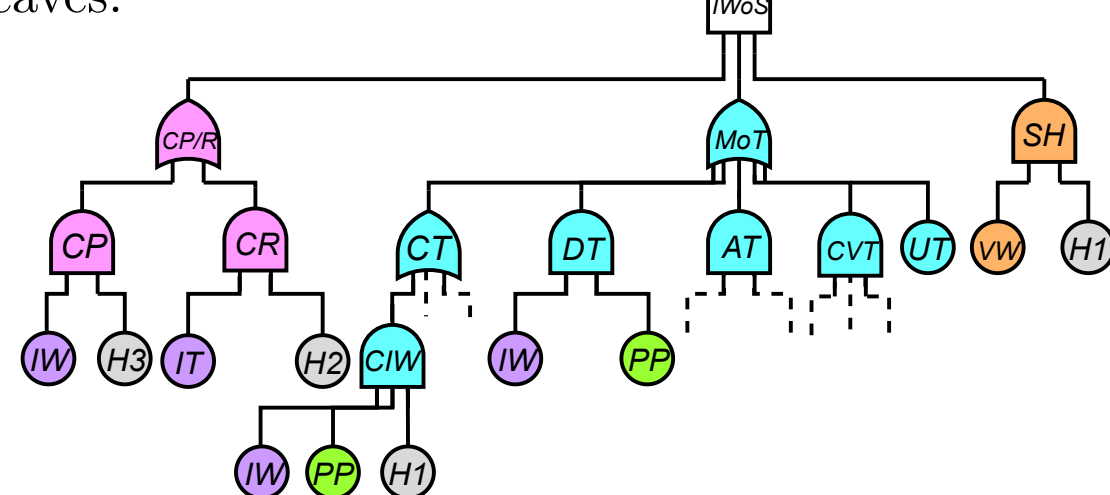

Figure 6: Tree for property 9 (excerpt from Fig. 2).

## VIII. Conclusions and Future Work

**Conclusions:** We presented *BFL*, a Boolean logic for FTs that enables the construction of complex queries that capture many relevant scenarios. *BFL* is a flexible yet powerful logic: as a testimony of this usefulness we showcased an application of *BFL* to a COVID19-related FT. Our logic can easily express properties and it can query FTs to account for diverse settings and what-if situations: specified properties can then be checked via the model checking algorithms we presented. Finally, we presented an algorithm to generate counterexamples which helps in narrowing the gap between the expressive power of *BFL* and real-life usability.

**Future work:** Our work opens several relevant perspectives for future research. First, it makes sense to extend *BFL* to model probabilities. Fault trees analysis often involves the computation of probabilistic dependability metrics, such as the system reliability, availability and mean time to failure, etc. A probabilistic fault tree logic will allow users to perform such quantitative analysis. Secondly, it is foreseeable to extend *BFL* in order to consider dynamic gates in FTs. To handle dynamic gates in dynamic FTs it would be very natural to have a logic that can express temporal properties, moving more in the direction of LTL [8] or CTL [7] or their timed variants TLTL [33] and TCTL [34]. Moreover, a step towards usability can be achieved through the development of a *Domain Specific Language* for *BFL*: in doing so, we would near the gap between the theoretical development of our logic and the need of practitioners for useful tools. Lastly, applying an implementation of this logic to an industrial case study could further propel the development of *BFL* by providing hands-on feedback from engineers and risk managers acquainted with FTA.